\documentclass[letter,11pt]{article}
\pdfoutput=1 
\usepackage{jheppub} 
\usepackage{amsfonts}
\usepackage{comment}
\usepackage{xcolor}
\usepackage{times}
\usepackage{graphicx,tabularx}
\usepackage{amsmath, amssymb, amsthm, amsfonts}
\usepackage{multirow}
\usepackage{changepage}
\usepackage{amssymb}
\usepackage{graphicx}
\usepackage{slashed}
\usepackage{bbm}

\definecolor{green2}{rgb}{0,0.56,0.32}
\definecolor{purple2}{rgb}{0.74902, 0.12549, 0.760784}

\RequirePackage[normalem]{ulem}

\def\figureautorefname~#1\null{Fig.\,#1\null}
\def\tableautorefname~#1\null{Table\,#1\null}

\def\equationautorefname~#1\null{Eq.\,(#1)\null}

\title{{  Precision Higgs Couplings in Neutral Naturalness Models: an Effective Field Theory  Approach} }

\author[a]{Lucien Heurtier}
\author[b]{, Hao-Lin Li}
\author[a]{, Huayang Song}
\author[a]{, Shufang Su}
\author[c]{, Wei Su}
\author[b,d,e]{, Jiang-Hao Yu}

\affiliation[a]{Department of Physics, University of Arizona, Tucson, AZ 85721, USA}
\affiliation[b]{CAS Key Laboratory of Theoretical Physics, Institute of Theoretical Physics, \\Chinese Academy of Sciences, Beijing 100190, P. R. China}
\affiliation[c]{
ARC Centre of Excellence for Particle Physics at the Terascale, \\Department of Physics,University of Adelaide, South Australia 5005, Australia}
\affiliation[d]{School of Physical Sciences, University of Chinese Academy of Sciences, \\No.19A Yuquan Road, Beijing 100049, P.R. China}
\affiliation[e]{School of Fundamental Physics and Mathematical Sciences, Hangzhou Institute for Advanced Study, University of Chinese Academy of Sciences, Hangzhou 310024, China}

\emailAdd{heurtier@arizona.edu}
\emailAdd{lihaolin@itp.ac.cn}
\emailAdd{huayangs@email.arizona.edu}
\emailAdd{shufang@email.arizona.edu}
\emailAdd{wei.su@adelaide.edu.au}
\emailAdd{jhyu@itp.ac.cn}

\preprint{ADP-20-20/T1130}

\abstract{ The Higgs sector in neutral naturalness models provides a portal to the hidden sectors, and thus measurements of Higgs couplings at current and future colliders play a central role in  constraining  the parameter space of the model.   We investigate a class of neutral naturalness models, in which the Higgs boson is a pseudo-Goldstone boson from the universal ${\rm SO}(N)/{\rm SO}(N-1)$ coset structure.  Integrating out the radial mode from the spontaneous global symmetry breaking, we obtain various  dimension-six operators in the Standard Model effective field theory, and calculate the low energy Higgs effective potential with radiative corrections included.  
We perform a $\chi^2$ fit to the Higgs coupling precision measurements at current and future colliders and show that the new physics scale could be explored up to 2.7 (2.8) TeV without (with) the Higgs invisible decay channels at future Higgs factories.}

\begin{document} 
\maketitle
 \flushbottom

\newpage

\section{Introduction}
\label{sec:intro}

 The naturalness of the energy scales which are present in the Standard Model (SM) and the stability of the Higgs mass under quantum fluctuations have sourced extremely prolific studies in the past decades. The extraordinary theoretical elegance of supersymmetric
theories and composite Higgs models  has led to many efforts to investigate the possible   presence of predicted colored top partners at the hadron colliders. However,  
no experimental evidence has been found for colored top partners with masses up to a few TeV~~\cite{Aaboud:2017ayj,ATLAS:2018iwl,Sirunyan:2016jpr,Khachatryan:2017rhw}.  Therefore, neutral naturalness models~\cite{Craig:2015pha, Craig:2014aea, Chacko:2005pe,Burdman:2006tz,Poland:2008ev,Cai:2008au,Geller:2014kta,Barbieri:2015lqa,Low:2015nqa,Serra:2017poj,Csaki:2017jby,Dillon:2018wye,Cohen:2018mgv,Cheng:2018gvu,Xu:2018ofw}   in which top partners do  not carry any SM Quantum Chromodynamics (QCD) color are proposed as alternative scenarios to
alleviate this tension.
 
In a class of neutral naturalness models, such as twin Higgs~\cite{Chacko:2005pe}, minimal neutral naturalness~\cite{Xu:2018ofw}, brother/trigeometric Higgs~\cite{Csaki:2017jby,Serra:2017poj}, 
the SM Higgs boson is identified as a pseudo Nambu-Goldstone Boson (pNGB) from a spontaneous global symmetry breaking, and the corresponding top partner that help to solve the little hierarchy problem carries a hidden QCD charges. Therefore, the conventional smoking-gun signature of  a top partner at the LHC does not apply to this type of models.   On the other hand, the hidden QCD sector interacts with the visible sector via the  exchange of the Higgs boson, therefore neutral naturalness models generically stand as prototypes of a Higgs portal to new physics.

As a result, three important effects can be used to probe the parameter space in such models.
First, there will be modifications of the Higgs couplings to the SM particles that mainly originate from the pseudo-Goldstone nature of the Higgs boson, which can be universally parameterized by the ratio $v/f$ characterizing the misalignment between the electroweak scale $v$ and the spontaneous global symmetry breaking scale $f$~\cite{Alonso:2016btr,Alonso:2016oah}. 
Measuring the deviation of the Higgs couplings from their SM values precisely provides an indirect probe of the new physics scale $f$. 
Second, depending on the ultra-violet (UV) dynamics, there could exist a radial mode from the spontaneous global symmetry breaking, identified as a heavy scalar boson.   The presence of such a heavy-scalar boson at a scale of order ${\cal O}(\mathrm{TeV})$ could be  tested at current and future particle colliders~\cite{Chacko:2017xpd,Ahmed:2017psb}. 
Third, after the electroweak symmetry breaking (EWSB), the mixing between   this heavy scalar and the SM Higgs boson also induces interactions between the SM Higgs boson and the mirror sector particles, such as, mirror botton, mirror glue-balls, etc. This will open several channels of the Higgs invisible decay, which can also be probed in the future Higgs precision measurement experiments.

After the discovery of a SM-like Higgs boson of 125 GeV at the LHC~\cite{Aad:2012tfa,Chatrchyan:2012xdj}, several proposals of future Higgs factories have been discussed around the world, including the Future Circular Collider (FCC-ee)~\cite{Gomez-Ceballos:2013zzn,fccpara,fccplan,Abada:2019zxq, Abada:2019lih} at CERN, Circular Electron Positron Collider (CEPC)~\cite{CEPC-SPPCStudyGroup:2015csa,CEPCStudyGroup:2018ghi} in China, as well as the  International Linear Collider (ILC)~\cite{Baer:2013cma,Bambade:2019fyw,Fujii:2019zll} in Japan.  With around $10^6$ Higgs produced, precision measurements of Higgs mass and couplings are available at percent level or better.  For example, $ZZh$ coupling can be measured at sub-percent level of about $0.2\%$.  
Therefore these precision measurements can be used as indirect probes of the neutral naturalness models mentioned above.

In this paper we aim to study rigorously the quantum corrections  to the SM Higgs potential and interactions of the   Higgs with SM and mirror particles due to the presence of the heavy scalar sector, using an effective field theory (EFT) approach. First we classify several benchmark models: the twin Higgs~\cite{Chacko:2005pe}, minimal neutral naturalness~\cite{Xu:2018ofw}, brother/trigeometric Higgs~\cite{Csaki:2017jby,Serra:2017poj}, into a universal ${\rm SO}(N)/{\rm SO}(N-1)$ coset framework based on both   linear and nonlinear descriptions.   We systematically write down the pseudo-Goldstone Higgs   Lagrangian with the radial mode included. By integrating out the heavy radial mode of the theory, which naturally introduces a UV scale $m_\sigma$, 
we derive the effective Lagrangian for the pseudo-Goldstone Higgs boson, and then match this Lagrangian to  a set of dim-6 operators in the SM EFT framework. 
 
Given the scalar potential at the UV scale, we derive the RG-improved effective Higgs potential generated by radiative corrections and obtain the radiatively corrected Higgs mass.  Two of the model parameters can be fixed by requiring the Higgs mass $m_h=125$ GeV and the Higgs vacuum expectation value (vev) $v=246$ GeV. 
After obtaining all the Higgs couplings at electroweak scale, we perform a global fit of the Higgs precision measurements as well as the invisible Higgs decay width for current and future colliders, which could provide   constraints of the new physics scale $f$ and mass of the radial mode $m_{\sigma}$.

The paper is organized as follows. In Sec.~\ref{sec:model},  we introduce the benchmark models and discuss the implications of different choices of cosets and representations in terms of particle  contents in the mirror sector. In Sec.~\ref{sec:EFT},  we present our general procedure of deriving an effective field theory for the Higgs scalar and its interactions with mirror and SM particles at UV scale. In Sec.~\ref{sec:RGp}, we discuss the renormalization group (RG) running of the Higgs potential parameters and derive the Higgs couplings at the electroweak scale, as well as the relevant invisible Higgs decay width. In Sec.~\ref{sec:globalfit}, we present the global fit results and discuss their implications.  In Sec.~\ref{sec:conclusion}, we conclude.  In appendix~\ref{appendixA}, we present several useful formulae for the universal ${\rm SO}(N)/{\rm SO}(N-1)$ coset description and for the ${\rm SU}(4)/{\rm SU}(3)$ coset in the content of twin Higgs model, in the non-linear representations.

\section{The model Setup}\label{sec:model}

 The motivation behind neutral naturalness models is to solve the little hierarchy problem or the LEP paradox~\cite{Barbieri:2000gf}, namely the hierarchy between the weak scale $\sim$100 GeV and the scale of new physics around 5$-$10 TeV indicated by the electroweak precision   data. The SM Higgs boson is identified as the pNGB associated with the spontaneous breaking of a global symmetry. In the absence of an explicit breaking of such symmetry, the SM Higgs boson is   massless at all orders in the perturbation theory.
However, interactions which explicitly break the global symmetry, such as the top Yukawa coupling, do  generate  a quadratically divergent  mass term for the Higgs. Such divergences  are exactly cancelled at the one-loop order by adding to the model, for instance, a top partner $\tilde t$ and a global $\mathbb Z_2$ symmetry of the Lagrangian under the exchange $t \leftrightarrow \tilde t$.
Indeed, the quadratic contributions to the Higgs potential from the top quark and top partner respect the global symmetry due to the imposed discrete symmetry, and  does not contribute to the Higgs mass.  The Higgs mass  is therefore only sensitive to the cutoff scale logarithmically at one-loop, alleviating the little Hierarchy problem.  In the meantime, contrary to the composite Higgs model and the supersymmetric model, the top partners in the neutral naturalness models introduced via certain $\mathbb Z_2$ symmetry are usually charged under a hidden SU(3) instead of the SM QCD gauge group, which also helps to evade the direct search constraints at the LHC.
 
There are different neutral naturalness models, depending on the choice of different global symmetry and coset structures. In particular, we consider the twin Higgs model~\cite{Chacko:2005pe}  as well as neutral naturalness models based on the cosets ${\rm SO}(5)/{\rm SO}(4)$\cite{Xu:2018ofw} and ${\rm SO}(6)/{\rm SO}(5)$~\cite{Csaki:2017jby,Serra:2017poj}.   The twin Higgs model~\cite{Chacko:2005pe}  is the prime and first example of neutral naturalness models, which introduces mirror copies of SM particles, charged under the mirror SM gauge groups only. However, this setup encounters a tension with cosmological observations, because the predicted light mirror fermions and mirror photon from the $Z_2$ symmetry provide too large contributions to the correction of the total radiation density characterized by $\Delta N_{\rm eff}$~\cite{Chacko:2016hvu, Csaki:2017spo}.     Thus neutral naturalness models based on the cosets ${\rm SO}(5)/{\rm SO}(4)$\cite{Xu:2018ofw} (minimum neutral naturalness)  and ${\rm SO}(6)/{\rm SO}(5)$~\cite{Csaki:2017jby,Serra:2017poj} (brother/trigeometric Higgs)  are introduced in order to solve the little hierarchy problem and as a consequence,   cosmological tensions can be avoided by reducing the amount of fermions and gauge bosons in the mirror sector.

We express the Lagrangian of the  scalar sector of these models as
\begin{eqnarray}\label{eq:scalarpotential}
{\cal L}_{S}=(D_{\mu}{\cal H})^\dagger(D^{\mu}{\cal H})-V_{sym}({\cal H})-V_{break}({\cal H}),
\end{eqnarray}
where $\cal H$ is the multiplet of the corresponding unbroken global symmetry containing the SM Higgs field.
{The potentials $V_{sym}$ and $V_{break}$ respectively preserve and explicitly break the global symmetry and can be defined as
}
\begin{eqnarray}
{V}_{sym}&=&-\mu^2 |{\cal H}|^2+\lambda|{\cal H}|^4\,,\\
{V}_{break}&=& +{\cal H}^\dagger \mathbf m^2 {\cal H}+ \left|{\cal H}^\dagger \boldsymbol \delta {\cal H}\right|^2.
\end{eqnarray}
The terms in $V_{break}$ contribute to the mass of the pseudo Nambu-Goldstone Boson.  The parameter matrix $\boldsymbol \delta$ is usually generated by loop effects from the gauge and Yukawa interactions that explicitly break the global symmetry, while $\bold m^2$ is a soft breaking term introduced by hand which -- in addition to breaking the global symmetry -- breaks the possible discrete symmetry to generate the misalignment between the SM Higgs vev and the spontaneous global symmetry breaking vev.    The smallness of $\boldsymbol \delta$ is guaranteed by their loop origin, while the smallness of $\bold m^2$ is technically natural in terms of the discrete symmetry.   
According to the Haag theorem, the scalar field ${\cal H}$ could be realized linearly or nonlinearly while keeping the physics the same.  Depending on the linear or nonlinear realization of the scalar fields, expressions for $\cal H$ , $\bold m^2$ and $\bold \delta$ are different, as we discuss below.

\subsection{Linear realization} 
In the linear realization, we can express $\cal H$ as:
\begin{eqnarray}
{\cal H}=\begin{pmatrix}
H_A \\
S
\end{pmatrix}\,,
\end{eqnarray}
where $H_A$ is the Higgs doublet that couples to SM fermions, while the field $S$ takes the following form depending on the models considered:
\begin{equation}
    S=\left\{
                \begin{array}{ll}
                  H_B \quad &{\rm twin\ Higgs}\,,\\
                  s/\sqrt{2}\quad &{\rm SO}(5)/{\rm SO}(4)\,,\\
                  (s+ia)/\sqrt{2}\quad &{\rm SO}(6)/{\rm SO}(5)\,.
                \end{array}
              \right.
              \label{eq:S}
\end{equation}
The matrices $\bold m^2$ and $\boldsymbol \delta$ under the linear parameterization have the following forms:
\begin{eqnarray}\label{eq:parameter_matrices}
\bold m^2=
\begin{pmatrix}
m^2 {1}_{n\times n}& 0\\
0 & -m^2 {1}_{p\times p}
\end{pmatrix}\,,
\quad
\boldsymbol \delta=
\begin{pmatrix}
\sqrt{\delta} {1}_{n\times n}& 0\\
0 & -i\sqrt{\delta}{1}_{p\times p}
\end{pmatrix}\,
\label{eq:mats}
\end{eqnarray}
where the integers $n$ and $p$ depend on the different coset structures and are summarized in Table~\ref{tab:np_models}.

\begin{table}\center
\begin{tabular}{c|cc|cc}
\hline 
\hline
 \multirow{2}{*}{coset} & \multicolumn{2}{c|}{linear} &\multicolumn{2}{c}{non-linear} \\ \cline{2-5}
&  n & p  & n & p  \\
 \hline 
twin\ Higgs &2 & 2 &  4 & 4     \\
${\rm SO}(5)/{\rm SO}(4)$  & 2 & 1 & 4 & 1    \\
${\rm SO}(6)/{\rm SO}(5)$ & 2 & 1 & 4 & 2  \\
\hline 
\hline
\end{tabular}
\caption{$n$ and $p$ for different neutral naturalness models in the linear and non-linear realization.}
\label{tab:np_models}
\end{table}

The covariant derivative can be expressed as
\begin{eqnarray}
D_{\mu}&=&\begin{pmatrix}
D_{\mu}^A & 0_{n\times p}\\
0_{p\times n} & D_{\mu}^B
\end{pmatrix},
\label{eq:Ds}
\end{eqnarray}
where $D_{\mu}^A$ is the ordinary covariant derivative of the SM gauge group, while $D_\mu^B$ is the corresponding covariant derivative  of the mirror gauge sector for the twin Higgs model and ${\rm SO}(6)/{\rm SO}(5)$ model, and is just {the} ordinary derivative $\partial_{\mu}$ for the ${\rm SO}(5)/{\rm SO}(4)$ models~\footnote{Here 
for simplicity, in the ${\rm SO}(6)/{\rm SO}(5)$ setup, we gauge an extra U(1) subgroup and thus the additional Goldstone boson is eaten.  We assume that this gauge boson is heavier than the Higgs boson and does not contribute to an additional invisible decay channel.  Another choice is keeping the additional Goldstone boson in the low energy spectrum. It could serve as a dark matter candidate.}.

In order to cancel the quadratic divergence arising from the radiative correction to the Higgs mass from the SM top loop, one typically introduces top partners through the Lagrangian~\footnote{In the ${\rm SO}(5)/{\rm SO}(4)$ coset, due to the kinetic normalization of the fifth component in the ${\rm SO}(5)$ representation, we need to either introduce a $\sqrt{2}$ factor in the top Yukawa sector or introduce additional doublet top partner to correctly cancel the quadratic divergence in the Higgs potential. Here we adopt the former, given $S=s/\sqrt{2}$ in Eq.~(\ref{eq:S}).  For the case with additional doublet top partner we refer readers to Ref.~\cite{Xu:2018ofw}. }
\begin{equation}
    {\cal L}_F= 
                  \lambda_t\bar{Q}_LH^c_A t_R + \left\{
                 \begin{array}{ll}\tilde{\lambda}_t\bar{\tilde{Q}}_L H_B^c \tilde{t}_R+h.c. \ \ & {\rm twin\ Higgs}\\
					 \tilde{\lambda}_t \bar{\tilde{t}}_L S \tilde{t}_R+h.c.\ \  &{\rm SO}(5)/{\rm SO}(4)\ {\rm and}\  {\rm SO}(6)/{\rm SO}(5)
					 \end{array}\right. , 
\end{equation}
for $H^c\equiv\epsilon H^*$ with $\epsilon$ being the $2\times 2$ Levi-Civita tensor.   In what follows,  we will focus on the case   $\lambda_t=\tilde \lambda_t$ at the scale where we   define our EFTs, which can be motivated by the presence of an accidental $\mathbb Z_2$ symmetry \cite{Chacko:2005pe} or a trigonometric parity \cite{Csaki:2017jby}.

\subsection{Nonlinear parameterization}
In order to see the explicit  pNGB nature of the physical Higgs, it is convenient to parameterize the multiplet $\mathcal H$ nonlinearly in the fundamental representation of the corresponding global symmetry ${\rm SO}(N)$:
\begin{eqnarray}
{\cal H}= \left(f+\frac{\sigma}{\sqrt{2}}\right)e^{i\frac{\sqrt{2}\Pi_a T^{\hat{a}}}{f}}\Phi =  \left(f+\frac{\sigma}{\sqrt{2}}\right) {\mathcal U}\Phi,
\end{eqnarray}
where we introduced the broken generators   denoted by $T^{\hat{a}}$ (as opposed to the unbroken generators denoted by $T^a$).   We chose the vacuum direction to lie along the last component of the multiplet, i.e. $\Phi=(0,\ldots,0,1)^T$. The scale $f$ is the vev related to the spontaneous global symmetry breaking, and $\sigma$ is the radial mode. The goldstone matrix ${\mathcal U}$ can be written as
\begin{eqnarray}
{\mathcal U}= \left(
\begin{array}{c|cc}
    \multicolumn{1}{c|}{\multirow{1}{*}{${1}_{(N-1)\times (N-1)}-\left(1-\cos{\frac{|\Pi|}{f}}\right)\frac{\Pi_i\Pi^\dagger_j}{|\Pi|^2}$}} &   \multirow{2}{*}{$\frac{\Pi_i}{|\Pi|}\sin{\frac{|\Pi|}{f}}$} \\
    &  & \\
    \hline
    \multicolumn{1}{c|}{  
    \begin{matrix}
  		-\frac{\Pi^\dagger_j}{|\Pi|}\sin{\frac{|\Pi|}{f}}
  \end{matrix}}   &  \cos{\frac{|\Pi|}{f}} 
\end{array}
\right)\,,
\end{eqnarray}
where $|\Pi|=\sqrt{\sum_{i=1}^{N-1}\Pi_i^2}$ and ${N-1}$ is the number of broken generators.  $N=8,5,6$ are the dimensions of the fundamental representation of the corresponding global symmetry groups of the twin Higgs, ${\rm SO}(5)/{\rm SO}(4)$ and ${\rm SO}(6)/{\rm SO}(5)$ models,  respectively.   Given these   notations, the field ${\cal H}$ can   be expressed as
\begin{eqnarray}
{\cal H}=\left(f+\frac{\sigma}{\sqrt{2}}\right)\times\left(\frac{\Pi_1}{|\Pi|}\sin\frac{|\Pi|}{f}\,,\ldots, \frac{\Pi_i}{|\Pi|}\sin\frac{|\Pi|}{f}\,,\hdots,\frac{\Pi_{N-1}}{|\Pi|}\sin\frac{|\Pi|}{f}\,,\ \cos \frac{|\Pi|}{f} \right)^T.
\end{eqnarray}
 The first four Goldstones $\vec{\Pi}=(\Pi_1,\Pi_2,\Pi_3,\Pi_4)$   can be grouped into an complex doublet representation of ${\rm SU(2)}_L$, which is the ordinary form of SM Higgs doublet $H$:
\begin{eqnarray}
H\equiv
\left(
\begin{array}{c}
\Pi_2+i\Pi_1,\\
\Pi_4-i\Pi_3
\end{array}
\right).
\label{eq:PitoH}
\end{eqnarray} 
The remaining Goldstones $\left\{\Pi_i\right\}_{i=5,\hdots,N-1}$ in the twin Higgs model, and ${\rm SO}(6)/{\rm SO}(5)$ model are eaten by the corresponding gauge bosons and can be set to be zero under the unitary gauge.    Note that for the twin Higgs model, in Ref.~\cite{Ahmed:2017psb}, the authors wrote down the Goldstone matrix based on the fundamental representation of ${\rm SU}(4)/{\rm SU}(3)$ which is complex, while our expressions are based on the fundamental representation of ${\rm SO}(8)/{\rm SO}(7)$ which is real. The two cosets are locally isomorphic to each other at renormalizable level.    We list the forms of ${\cal H}$ for both cases in the appendix~\ref{appendixA} as a comparison.

Two matrices  $\bold m^2$ and ${\boldsymbol \delta}$ and the corresponding derivative $D_\mu$ have the same forms as in Eq.~(\ref{eq:mats}) and ~(\ref{eq:Ds}) respectively, with the values of $n$ and $p$ depending on the different coset structures listed also in Table~\ref{tab:np_models}.  
The concrete forms of $D^A_{\mu}$, and $D^B_{\mu}$ for different coset structures are summarized in Table~\ref{tab:der_models} in appendix~\ref{appendixA}.
 
The Yukawa coupling in the nonlinear representation depends on the embedding of the top quark in the representation of the global symmetry.   In general the relevant top sector Yukawa couplings can   be written as
\begin{equation}\label{eq:UVinteractions}
\mathcal L_F\supset \left(f+\frac{\sigma }{\sqrt{2}}\right)\left[\lambda_{t}\bar{\Psi}_{L}{\cal U}\Psi_{R}+\tilde{\lambda}_t\bar{\tilde{\Psi}}_{L}{\cal U}\tilde{\Psi}_{R}\right]+h.c.,
\end{equation}
where $\Psi_{L,R}$ and $\tilde{\Psi}_{L,R}$ are incomplete multiplets of the corresponding fundamental representations of the global symmetry  where the SM left handed quark doublet and the corresponding right handed fields and their partners  are embedded:  
 \begin{eqnarray}
 \Psi_{L}&=& \begin{cases} ( i b_L, - b_L,  i t_L,  t_L,0,0,0,0)^T\quad \textrm{twin\ Higgs}\,,\\
 (i b_L, - b_L,  i t_L,  t_L,0,0)^T\quad {\rm SO}(6)/{\rm SO}(5)\,, \nonumber\\
 ( i b_L, - b_L,  i t_L,  t_L, 0)^T\quad {\rm SO}(5)/{\rm SO}(4)\,,
\end{cases}\, \nonumber\\
\Psi_{R}&=& (0,0,0,...,t_R)^T\,,\nonumber\\
\tilde{\Psi}_{L}&=& \begin{cases} (0,0,0, 0, i\tilde{b}_L, -\tilde{b}_L, i\tilde{t}_L, \tilde{t}_L)^T\,, \quad \textrm{twin Higgs}\,,\\
(0,0,0, 0, i\tilde{t}_L, \tilde{t}_L)^T\,, \quad  {\rm SO}(6)/{\rm SO}(5)\,,\\
(0,0,0, 0, \sqrt{2}\tilde{t}_L)^T\,, \quad {\rm SO}(5)/{\rm SO}(4)\,,
\end{cases}
\nonumber\\
\tilde{\Psi}_{R}&=& (0,0,0,...,\tilde{t}_R)^T.
\end{eqnarray}
Readers can refer to Ref.~\cite{Xu:2018ofw,Csaki:2017jby,Li:2019ghf,Barbieri:2015lqa} for more details and different embeddings.
For the twin Higgs model, in order to avoid the anomaly, we also introduce $\tilde{b}_R$, as well as twin taus and twin neutrino\footnote{The Majorana mass of the twin neutrino is set to be larger than MeV to avoid cosmological constraints. For detailed discussions on how to lift the the twin neutrino masses, we refer readers to Ref.~\cite{Feng:2020urb}.  The existence of mirror neutrino does not affect the Higgs phenomenology since the corresponding coupling is small. } following the ``Fraternal Twin Higgs" setup~\cite{Craig:2015xla}.

\section{Effective Field Theory in the Non-Linear Representation}\label{sec:EFT}
In this section, we present the method we use in order to obtain the effective Lagrangian for the SM Higgs doublet $H$ at UV scale by integrating out the radial mode $\sigma$ and derive all the couplings of the physical Higgs boson to the SM and mirror particles. Such couplings will allow us to   derive the radiative corrections to the parameters in the Higgs potential arising from the renormalization group running.  Note that in our analyses below, we use the nonlinear realization for pseudo Goldstone Higgs bosons.

The covariant derivative term in the Lagrangian can be simplified as
\begin{eqnarray}\label{eq:UVkinetic}
(D_{\mu}{\cal H})^\dagger(D^{\mu}{\cal H})&=&\left(f+\frac{\sigma}{\sqrt{2}}\right)^2\frac{1}{2}{\rm Tr}[d_\mu d^\mu]+\frac{1}{2}\partial_\mu \sigma \partial^\mu \sigma\,.
\end{eqnarray}
The vector $d_\mu$ is defined by the decomposition of Maurer-Cartan form constructed with the Goldstone matrix ${\cal U}$~\cite{Coleman:1969sm,Callan:1969sn}  
\begin{eqnarray}
-i {\cal U}^\dagger D_{\mu} {\cal U} = d_{\mu}^{\hat{a}}T^{\hat{a}}+E_{\mu}^{a}T^a \equiv d_\mu + E_\mu.
\end{eqnarray} 
Working in the unitary gauge and using Eq.~\eqref{eq:PitoH},  one finds that,   
\begin{align}
\frac{1}{2}{\rm Tr}[d_\mu d^\mu]=&
\frac{1}{|H|^2}\sin^2\frac{|H|}{f}|D^A_{\mu}H|^2+\frac{1}{4|H|^4}\left(\frac{|H|^2}{f^2}-\sin^2\frac{|H|}{f}\right)(\partial_\mu|H|^2)^2 + \ldots
\end{align}
where $|H|=\sqrt{\sum_{i=1}^4\Pi_i^2}$ and ``$\ldots$" are the terms involving mirror gauge bosons.   
In the case of the twin Higgs model, those terms can be written as
\begin{equation}
\ldots=\cos^2\frac{|H|}{f}\left(\tilde{g}^2(|\tilde{W}^1_\mu|^2+|\tilde{W}^2_\mu|^2)+|\tilde{g}^{'}\tilde{B}_\mu-\tilde{g}\tilde{W}_\mu^3|^2\right),
\end{equation}
where $\tilde W^a_\mu$ and $\tilde B_\mu$ are the corresponding twin gauge fields as in the original twin Higgs model~\cite{Chacko:2005pe}.  
This Lagrangian gives rise to a massless twin photon in addition to the heavy twin $W$ and twin $Z$ bosons.
To avoid the cosmological $\Delta N_{\rm eff}$ problem caused by the massless twin photon, one either introduces a kinetic mixing between the SM hypercharge and the twin hypercharge gauge fields, or
simply removes the twin hypercharge gauge boson degree of freedom.
For the coset ${\rm SO}(6)/{\rm SO}(5)$, we choose to gauge the mirror U(1) to absorb the residue single Goldstone boson, which has been investigated in detail in Ref.~\cite{Qi:2019ocx}. In the case of the minimal coset  ${\rm SO}(5)/{\rm SO}(4)$, there is no additional Goldstone boson other than the SM ones, and thus 
there is no need to introduce mirror gauge bosons.  Overall, in all cases we choose to introduce mirror gauge bosons to absorb  additional Goldstone bosons other than the SM Higgs sector.

\begin{figure}
    \centering
    \includegraphics[width=0.7\linewidth]{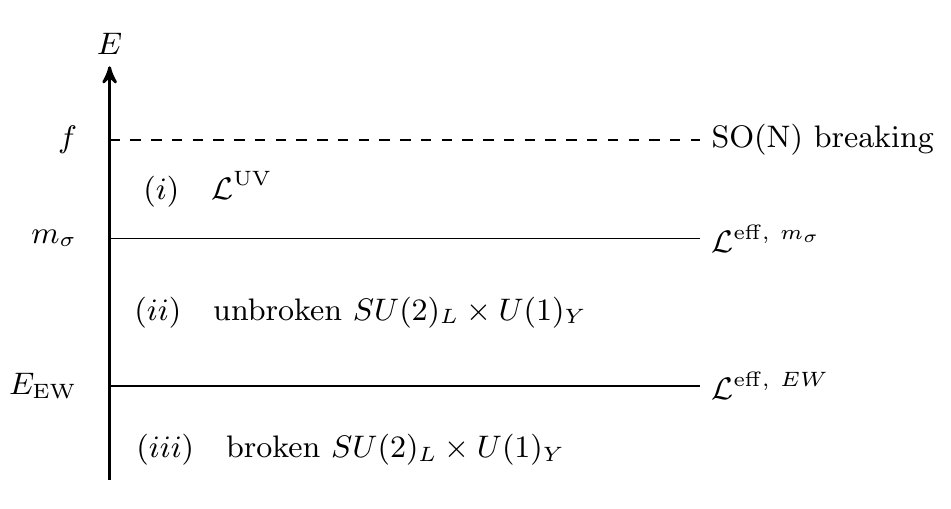}
    \caption{\label{fig:energyregimes}Different energy regimes considered throughout the paper.}
\end{figure}
Moreover, the scalar potential of Eq.~\eqref{eq:scalarpotential} can be written 
\begin{equation}\label{eq:UVpot}
V_{sym}+V_{break}=\left(f+\frac{\sigma}{\sqrt{2}}\right)^2\left(-\mu^2-m^2\cos\frac{2|{H}|}{f}\right)+\left(f+\frac{\sigma}{\sqrt{2}}\right)^4\left(\lambda+\delta-\frac{\delta}{2}\sin^2\frac{2|{H}|}{f}\right)\,.
\end{equation}

We can now distinguish three different energy regimes:   $(i)$ In the UV limit, the scalar theory we are considering consists of two scalar fields $H$ and $\sigma$ and four real parameters $\mu$, $f$, $m$ and $\delta$.  This typically is the case as long as both scalar fields are dynamical degrees of freedom, {\em id est} when the typical energy of the physical process considered is above the threshold $E\gtrsim m_\sigma\gg m_h$ where $m_\sigma$ and $m_h$  denote the masses of the radial mode and the SM Higgs boson, respectively. $(ii)$ Below $m_\sigma$, but above the electroweak scale, the scalar sector can be described by an effective field theory for the SM Higgs doublet, in which the scalar $\sigma$ has been integrated out. $(iii)$ As energies go under the electroweak scale $E<E_{EW}$, the effective Higgs potential undergoes spontaneous symmetry breaking, and the Higgs boson acquires a vev $\langle H\rangle\not = 0$. In the previous discussion we have derived the Lagrangian in the UV regime and obtained an explicit form for the scalar gauge kinetic terms as in Eq.~\eqref{eq:UVkinetic}, scalar potential as in Eq.~\eqref{eq:UVpot} and fermionic interactions as in Eq.~\eqref{eq:UVinteractions}. In what follows we will describe how to obtain the effective field theory for the physical Higgs boson at the scale $m_\sigma$ after integrating out the radial mode. In the next section we will then derive how the effective Lagrangian gets modified when running the theory from the scale $m_\sigma$ down to the electroweak scale. Then, after performing the spontaneous EWSB, we will finally end up with the physical effective Lagrangian for the Higgs boson at low energy.

At the UV scale, however, after spontaneous breaking of the global symmetry, the heavy scalar $\sigma$ stabilizes to the origin and we can define the mass of the $\sigma$ field in the phase where $\langle H \rangle =0$ as\footnote{Note that the physical mass of the heavy scalar as defined in the true vacuum should be calculated after electroweak symmetry breaking. However, in the limit of $\langle H \rangle/f\to 0$,  both quantities are equal at leading order, which we use for simplicity to parameterize our effective field theory.   }     
\begin{eqnarray}
m_\sigma^2 = -\left.\frac{\partial^2 \cal L_S}{\partial \sigma^2}\right|_{\sigma , H=0} =6f^2(\lambda+\delta)-\mu^2-m^2,
\end{eqnarray}
and one can solve the minimization condition for the field $\sigma$
\begin{eqnarray}\label{eq:tadpole}
\left.\frac{\partial \cal L_S}{\partial \sigma}\right|_{\sigma , H=0} =0\quad\to\quad \mu^2+m^2=2f^2(\lambda+\delta).
\end{eqnarray}
Using such relations, we can write the Lagrangian in the form of~\cite{Henning:2014wua}   
\begin{equation} \label{eq:lagrangianform}
\mathcal{L}_S=\sigma \cdot B-\frac{1}{2}\sigma (\partial^2+m_{\sigma}^2+U)\sigma +\mathcal{O}(\sigma^3)\,,
\end{equation}
 where $U=U(H,{D}_{\mu}^A H)$ and $B=B(H,{D}_{\mu}^A H)$ are functions of the Higgs field and its derivative only, which can be read from the Lagrangian as
\begin{align}\label{eq:UB}
U=&-\Bigg\{\frac{1}{2}{\rm Tr}[d_\mu d^\mu] +m^2\left(\cos\frac{2|{H}|}{f}-1\right)+3f^2\delta\sin^2\frac{2|{H}|}{f}\Bigg\}\,, \\ 
B=&\sqrt{2}f\Bigg\{\frac{1}{2}{\rm Tr}[d_\mu d^\mu]+\mu^2 +m^2 \cos\frac{2|{H}|}{f}-2f^2\left(\lambda+\delta-\frac{\delta}{2}\sin^2\frac{2|{H}|}{f}\right)\Bigg\}\,.
\end{align}

To integrate out the radial mode $\sigma$ by solving the classical equation of motion and organize the effective Lagrangian in ${\cal O}(p)$ expansion, we need to investigate the $p$ order of $U$ and $B$, for $p$ being the typical momentum scale of a low energy process.  We used such scaling in our $\mathcal O(p)$ expansion.
Since the mass and vev for the SM Higgs boson are at the electroweak scale, the parameters $m$ and $\sqrt \delta f$   need to be set to relatively small values as compared to the scale $f$ and $m_\sigma$.  For this reason, we will  identify these parameters as being $\mathcal O(p)$. Given these power-counting rules, one can read from Eq.~(\ref{eq:UB}) the scaling of functions $U$ and $B$ as  
\begin{equation}
U\sim \mathcal{O}(p^2)\,,\quad B\sim \mathcal{O}(p^2)\,.
\end{equation}
Solving the equation of motion for $\sigma$, and truncating the theory up to $\mathcal O(p^3)$, one obtains the classical solution $\sigma_c$, which is
\begin{eqnarray}
\frac{\delta \mathcal L_S}{\delta \sigma}=0 \quad\longrightarrow\quad 
\sigma_c&=&\frac{B}{\partial^2+m_{\sigma}^2+U}\approx \frac{B}{m_{\sigma}^2}+\mathcal{O}(p^4)\,.
\end{eqnarray}
Plugging the solution $\sigma_c$ into the Lagrangian $\mathcal L_S$ and making use of the tadpole condition of Eq.~\eqref{eq:tadpole} and the definition of $m_\sigma$, we obtain the effective Lagrangian\footnote{{Note that we did not include loop corrections to the following Lagrangian since they all come at the order $\mathcal{O}(p^6)$ in addition to the loop-suppression factor.}} up to ${\cal O}(p^4)$
\begin{eqnarray}
{\cal L}^{{\rm eff},~m_\sigma}_{non-linear}
&=&\frac{f^2}{2}{\rm Tr}[d_\mu d^\mu]-2f^2m^2\sin^2\frac{|H|}{f}+\frac{\delta f^4}{2}\sin^2\frac{2|H|}{f}\nonumber \\
&+&\frac{f^2}{m_\sigma^2}\left(\frac{1}{2}{\rm Tr}[d_\mu d^\mu]-2m^2\sin^2\frac{|H|}{f}+\delta f^2\sin^2\frac{2|H|}{f}\right)^2  +{\cal O}(p^6)+const.
\label{eq:EFTHnonlinear}
\end{eqnarray}
 Expanding the trigonometry function above to obtain the terms in polynomial forms in $H$ and performing a transformation to the Warsaw basis (See e.g. the Eq.~(24) of Ref.~\cite{Corbett:2017ieo}),  we obtain the effective Lagrangian for the SM Higgs doublet as
\begin{eqnarray}
\mathcal{L}^{{\rm eff},~m_\sigma}_{S}&=&|{D}_\mu^A H|^2-V^{{\rm eff},m_\sigma}\nonumber \\
&=&|{D}_\mu^A H|^2+\mu_H^2|H|^2-\lambda_H|H|^4 +\frac{c_{H}}{2f^2}\mathcal{O}_{H}+\frac{c_{6}}{f^2}\mathcal{O}_{6}\,,\label{eq:lagEFTscalar}
\end{eqnarray}
for 
\begin{equation}
\quad \mathcal{O}_{H}\equiv (\partial_\mu|H|^2)^2\,,\qquad\mathcal{O}_{6}\equiv |H|^6\,.
\end{equation}
We find that
\begin{equation}\label{eq:higgsparam}
\mu_H^2=2 \delta  f^2-2 m^2\,, \ \  \ 
\lambda_H=2 \delta +\frac{4 m^4}{f^2 m_{\sigma}^2}-\frac{8 \delta  m^2}{m_{\sigma}^2}.
\end{equation}
The Wilson coefficient appearing in the dimension-six operators are given by
\begin{equation}
c_{H}=\frac{1}{2}+\frac{4 m^2}{m_{\sigma}^2}-\frac{8 \delta f^2}{m_{\sigma}^2}\,, \ \ \ 
c_6=\frac{16 m^2}{45 f^2}-\frac{16 \delta }{45}\,.
\end{equation}

The EFT Lagrangian in the Higgs sector in Eq.~\eqref{eq:lagEFTscalar} are defined at the matching scale $m_\sigma$, which serves as the starting point to derive the RG improved Higgs potential in Sec.~\ref{sec:RGp}.

From Eq.~\eqref{eq:EFTHnonlinear}, one can also obtain the interaction between $H$ and twin gauge bosons in the twin Higgs model up to ${\cal O}(p^2)$ as   
\begin{eqnarray}\label{eq:lagEFTgauge}
\mathcal{L}^{{\rm eff},~m_\sigma}_{mirror\ gauge}&\supset& -|H|^2\left(1+\frac{4m^2}{m_\sigma^2}-\frac{8f^2\delta}{m_\sigma^2}\right)\left[\tilde{g}^2(|\tilde{W}^1_\mu|^2+|\tilde{W}^2_\mu|^2)+|\tilde{g}^{'}\tilde{B}_\mu-\tilde{g}\tilde{W}_\mu^3|^2\right]\nonumber\\
&&-\frac{2|H|^2}{m_\sigma^2}\left[\tilde{g}^2(|\tilde{W}^1_\mu|^2+|\tilde{W}^2_\mu|^2)+|\tilde{g}^{'}\tilde{B}_\mu-\tilde{g}\tilde{W}_\mu^3|^2\right]^2\ ,
\label{eq:twin_gauge}
\end{eqnarray}
An expression for the ${\rm SO}(6)/{\rm SO}(5)$ coset with only mirror U(1) gauge boson can be written down in a similar manner.

In the fermionic sector,  the Lagrangian we start with at the UV scale for the top and mirror top sector is given by   
\begin{equation}
\mathcal L_F^{\rm UV}\supset\left(f+\frac{\sigma }{\sqrt{2}}\right) \left[\lambda_t\frac{\bar{Q}_L H^ct_R}{|H|} \sin \left(\frac{|H|}{f}\right)+\tilde{\lambda}_t\bar{\tilde{t}}_L\tilde{t}_R \cos \left(\frac{|H|}{f}\right)+h.c.\right]\,. 
\end{equation}
 Terms which involve bottom quark, leptons and their partners can be written down similarly in the twin Higgs model. 
After following the procedure described earlier, integrating out the radial mode $\sigma$,  and going into the Warsaw basis, the effective Lagrangian for the fermion sector is\footnote{Note that in Refs.~\cite{Greco:2016zaz, Contino:2017moj}, the couplings of the Higgs to   the fermionic sector are obtained by performing the transformation $H\to f\frac{H}{|H|}\sin(|H|/f)$ (up to a $\sqrt 2$ difference coming from a different normalization of the scale $f$), which is equivalent to the transformation we performed here.}
\begin{equation}\label{eq:lagEFTfermions}
\mathcal L_{F}^{{\rm eff},~m_\sigma}\supset\left[\lambda_t\bar{Q}_L H^ct_R+ \tilde{\lambda}_t\bar{\tilde{t}}_L\tilde{t}_R f\left(1-\frac{|H|^2}{2f^2}+\frac{|H|^2}{f^2}\left(\frac{4\delta f^2}{m_\sigma^2}-\frac{2m^2}{m_\sigma^2}\right)\right) +h.c.\right]\,.
\end{equation}

\section{RG improved Higgs potential and couplings}
\label{sec:RGp}
At lower energies,  the Higgs potential receives important contributions from loops involving heavy states, in particular from the top quark and its mirror partner.   In the Minimal Supersymmetric Standard Model (MSSM), such contributions have been thoroughly studied, and the calculation of the Higgs masses after renormalization of the EFT at the three-loop order is presently available in several public codes to a very good precision~\cite{Bahl:2018qog}. However, for neutral naturalness models, the dominant contributions at next-to-next-leading order (NNLO) of this effect have only been carried out in the framework of the Composite twin Higgs model in Ref.~\cite{Greco:2016zaz, Contino:2017moj}, in which the authors have shown that a significant fraction of the Higgs mass actually comes from the running of the effective potential from the UV scale down to the electroweak scale. In this section we aim to evaluate such corrections to the Higgs potential based on the approach of Refs.~\cite{Greco:2016zaz, Contino:2017moj}, as a function of the UV parameters introduced in the previous sections.

Using the results of Sec.~\ref{sec:EFT} and gathering Eq.~\eqref{eq:lagEFTscalar}, Eq.~\eqref{eq:lagEFTgauge}, and Eq.~\eqref{eq:lagEFTfermions}, the interaction Lagrangian which we use as our starting point at  the scale $m_\sigma$ in order to calculate the Higgs potential at the electroweak scale is   given by  
\begin{align}
    \mathcal{L}^{{\rm eff},~m_\sigma}=&\mathcal{L}^{{\rm eff},~m_\sigma}_{S}+\mathcal{L}^{{\rm eff},~m_\sigma}_{F}+\mathcal{L}^{{\rm eff},~m_\sigma}_{mirror\ gauge}\,.
    \label{eq:LEFTdef}
\end{align} 
in the twin Higgs model.  A similar term for $\mathcal{L}^{{\rm eff},~m_\sigma}_{mirror\ gauge}$ for ${\rm SO}(6)/{\rm SO}(5)$ coset can be used with only the mirror U(1) sector.  
The Wilson coefficients at the scale $m_\sigma$ can be read from Sec.~\ref{sec:EFT}, while the $\mathbb Z_2$ symmetry enforces   
\begin{align}
    \lambda_t(m_\sigma)=\tilde{\lambda}_{t}(m_\sigma),\quad g_S(m_\sigma)=\tilde{g}_{S}(m_\sigma)\quad\text{and}\quad g_2(m_\sigma)=\tilde{g}_{2}(m_\sigma)\,.
\end{align}
 In our calculation, $\lambda_t(m_\sigma)$, $g_S(m_\sigma)$ and $g_2(m_\sigma)$ are computed from $\lambda_t(m_t)$, $g_S(m_t)$ and $g_2(m_t)$ using two-loop and one-loop fixed-order formulae, respectively:
\begin{align}
    &\lambda_t(m_\sigma)=\lambda_t(m_t)\bigg[1-\bigg(\frac{g_S^2(m_t)}{4\pi^2}-\frac{9\lambda_t^2(m_t)}{64\pi^2}\bigg)\log\frac{m_{\sigma}^2}{m_t^2}+\frac{22g_S^4(m_t)}{(4\pi)^4}\log^2\frac{m_{\sigma}^2}{m_t^2}\bigg]\,, \\
    &g_S(m_\sigma)=g_S(m_t)\bigg[1-\frac{7g_S^2(m_t)}{32\pi^2}\log\frac{m_{\sigma}^2}{m_t^2}\bigg]\,, \\
    &g_2(m_\sigma)=g_2(m_t)\bigg[1-\frac{19}{6}\frac{g_2^2(m_t)}{32\pi^2}\log\frac{m_{\sigma}^2}{m_t^2}\bigg]\,.
\end{align}

\subsection{Running of the scalar potential}
The Higgs potential that will be discuss below refers to the one particle irreducible (1PI) effective potential of the Higgs for the EFT defined in Eq.~\eqref{eq:LEFTdef} after integrating out the radial mode $\sigma$. From the well-known background field method~\cite{Jackiw:1974cv}, the 1PI effective potential $V_{\rm 1PI}(\phi)$ for a theory defined by a Lagrangian ${\cal L}[\phi]$ can be calculated by the vacuum energy $V_{\rm 1PI,b}(\phi=0;\phi_b)$ for an auxiliary theory defined by the Lagrangian ${\cal L}_b[\phi;\phi_b]\equiv{\cal L}[\phi_b+\phi]-{\cal L}'[\phi_b]\phi$ with $\phi_b$ a non-dynamical background field and $-{\cal L}'[\phi_b]\phi$ eliminating the tadpole terms from the relation~\cite{Schwartz:2013pla}:
\begin{equation}
    V_{\rm 1PI}(\phi_b) = V_{\rm 1PI,b}(0;\phi_b).
\end{equation}
Therefore the RG evolution of $V_{\rm 1PI}(\phi_b)$ in the original theory is equivalent to the RG evolution of the vacuum energy in  the auxiliary theory.   At the one-loop order, it  is given by~\cite{Ford:1992mv,Contino:2017moj,Greco:2016zaz}:
\begin{equation}
    \frac{dV_{\rm 1PI}(\phi_b,t)}{dt} = \frac{dV_{\rm 1PI,b}(0;\phi_b,t)}{dt} = \sum_f \frac{g_f}{64\pi^2}M^4_f -\sum_b \frac{g_b}{64\pi^2}M^4_b,
\end{equation}
where $t=\log(m_{\sigma}^2/\mu^2)$ characterizes the renormalization scale $\mu$ of the theory.   The sum over $f$ and $b$ are the contributions from $\phi_b$ dependent mass of fermions and bosons in the auxiliary ${\cal L}_b[\phi;\phi_b]$ theory.  The parameters $g_f$ and $g_b$ stand for the number of degrees of freedom of the corresponding fields (for instance, the top quark has $g_t =4\times 3=12$).  
In what follows, we will denote the background Higgs field as $H_c$ and solve the RG evolution for different sectors.

The fermion contribution to the vacuum energy is dominated by  
\begin{align}\label{eq:Vf}
    \frac{dV_F(H_c,t)}{dt}=\frac{3}{16\pi^2}\Big[M_t^4(H_c,t)+M^4_{\tilde{t}}(H_c,t)\Big]\,,
\end{align}
 where $M_t(H_c,t)$ and $M_{\tilde{t}}(H_c,t)$ are the running masses of the top and mirror top in the background of the Higgs field $H_c$
 and can be expressed as
\begin{align}
    M_t(H_c, t)=\lambda_t(H_c, t)H_c,\quad
    M_{\tilde{t}}(H_c, t)=\tilde\lambda_{{t}}(H_c, t)f\Big[1-(1-\frac{8\delta f^2}{m_\sigma^2}+\frac{4m^2}{m_\sigma^2})\frac{H_c^2}{2f^2}\Big].
\end{align}
 The RG-equations for the Yukawa couplings can be found following Refs.~\cite{Greco:2016zaz, Contino:2017moj} and one can obtain the expression of the potential $V_F$   by integrating Eq.~\eqref{eq:Vf} with respect to $t$.   

As far as the contributions coming from the gauge-sector are concerned, we only take into account the leading logarithmic term,  which is given by \footnote{Note that we corrected here an error found in Eq.~(C.4) of Ref.~\cite{Contino:2017moj} in which the powers of 2 should be powers of 4}.   
\begin{align}
    V_{gauge}(H_c,t)=-\frac{3}{64\pi^2}\Big[2M^4_W(H_c)+M^4_Z(H_c)+3M^4_{\tilde{W}}(H_c)\Big]t\,.
\end{align}
Again, for ${\rm SO}(6)/{\rm SO}(5)$ and ${\rm SO}(5)/{\rm SO}(4)$ model, $M^4_{\tilde{W}}(H_c)$ term does not exist.

There are  also contributions from the scalar field itself since there is an explicit self-interaction potential for the Higgs from the UV theory. The leading logarithmic contribution is   
\begin{align}
    V_S(H_c,t)=-\frac{1}{64\pi^2}\Big[M_H(H_c)\Big]^4t=-\frac{1}{64\pi^2}\Big[-\frac{1}{2}\mu_H^2+\frac{3}{2}\lambda_H h_c^2-\frac{15c_6}{8f^2}h_c^4\Big]^2t\,.
\end{align}
Combining all three terms, the full improved effective potential is given by
\begin{equation}
V^{\rm full}(H_c,t)=V^{{\rm eff}, m_\sigma}(H_c)+V^{\rm RGE}(H_c,t)\,,
\end{equation}
where
\begin{equation}
V^{\rm RGE}(H_c,t)=V_S(H_c,t)+V_{gauge}(H_c,t)+V_F(H_c,t)\,,
\end{equation}
and $V^{{\rm eff}, m_\sigma}$ serving as a boundary condition at the matching scale $m_\sigma$ can be read from Eq.~\eqref{eq:LEFTdef}.  The potential $V^{\rm RGE}(H_c,t)$ can be written at the energy scale given by the parameter $t$ in terms of parameters valued at the scale $m_\sigma$\footnote{{Note that in our numerical calculation, we only involve the dominant NNLO contributions following Ref.~\cite{Contino:2017moj} which points out that Ref.~\cite{Greco:2016zaz} misses some additional contributions from the twin Goldstone bosons. Without a complete NNLO calculation, the loss of accuracy on the Higgs mass prediction results in large uncertainties on the UV parameters. However this is not expected to affect significantly  the general conclusions of our findings.}}.

 Finally, the loop correction to the field strength of the Higgs doublet renders the kinetic terms in the Eq.~\eqref{eq:lagEFTscalar} non-canonical:
\begin{eqnarray}
|D_\mu^A H_c|^2\to Z_{H_c}|D_\mu^A H_c|^2,\quad    Z_{H_c}=1+\frac{3\lambda_t^2}{(4\pi)^2}t\,\label{eq:Zhc},
\end{eqnarray}
where we include, similarly to Refs.~\cite{Greco:2016zaz, Contino:2017moj}, only the leading logarithmic running effect coming from top loops. The Higgs doublet field can therefore be normalized using the redefinition $H_c\to H/\sqrt{Z_{H_c}}$.

\subsection{Couplings of the Higgs to Fermions and Gauge Bosons}
In the previous subsection we discussed the correction to the Higgs potential from the running effect, we are now ready to derive the physical Higgs coupling to fermions and gauge bosons.
First, we write down EFT at the $m_t$ scale by changing the couplings in Eq.~\eqref{eq:twin_gauge} and Eq.~\eqref{eq:lagEFTfermions} to the corresponding ones at the $m_t$ scale.    We also  make a renormalization of the Higgs doublet $H_c\to H/\sqrt{Z_{H_c}}$   such that the kinetic term of the Higgs doublet field is canonically normalized:   
\begin{eqnarray}
\mathcal{L}^{{\rm eff},~\rm EW}_{S}&=&|D_\mu^A H|^2+V^{\rm full}(H/\sqrt{Z_{H_c}},t)\,,\\
\mathcal L_{F}^{{\rm eff},~\rm EW}&\supset& \lambda_t(H, t)\frac{\bar{Q}_L H^c t_R}{\sqrt{Z_{H_c}}} +\tilde\lambda_{{t}}(H, t)\bar{\tilde{t}}_L\tilde{t}_R f\left[1-\frac{|H|^2}{2f^2Z_{H_c}}\left(1-\frac{8\delta f^2}{m_\sigma^2}+\frac{4m^2}{m_\sigma^2}\right)\right] +h.c.\,. \nonumber\\
\end{eqnarray}
 Next, we make the replacement $H\to (h+v)/\sqrt 2$ to obtain the couplings of the physical Higgs $h$ after the electroweak symmetry breaking, with $v=246$ GeV.
One can read the mass of the fermions and their mirror partners   
\begin{eqnarray}\label{eq:MirrorMass}
{m_t}&=&\frac{\lambda_t v}{\sqrt{2Z_{H_c}}}\,,\quad m_{\tilde{t}}=\tilde\lambda_t\left[f-\frac{v^2}{2Z_{H_c}}\left(\frac{1}{2f}+\frac{2m^2}{fm_{\sigma}^2}-\frac{4\delta f}{m_{\sigma}^2}\right)\right]\,,
\end{eqnarray}
where the coupling $\lambda_t$ and $\tilde \lambda_t$ are evaluated at the electroweak scale. 
After normalizing the kinetic term of the physical Higgs boson $h$ with the function
\begin{equation}
    Z_h=1+\frac{c_H}{Z_{H_c}^2}\frac{v^2}{f^2}\,,
\end{equation}
the Yukawa couplings become 
\begin{equation}\label{eq:YukawaCoupling}
\mathcal L_{htt}\supset \tilde\lambda_{t}^{\rm eff}h\bar{\tilde t}_L\tilde{t}_R + \lambda_{t}^{\rm eff} h\bar{t}_L t_R+h.c.\,, \end{equation}
with\begin{equation}
\tilde\lambda_{t}^{\rm eff}\equiv  -\frac{\tilde \lambda_t v}{2Z_{H_c}\sqrt{Z_h}} \left[\frac{1}{f}+\frac{4m^2}{fm_{\sigma}^2}-\frac{8\delta f}{m_{\sigma}^2}\right]\,,\text{ and ~~}\lambda_{t}^{\rm eff}\equiv \frac{\lambda_t}{\sqrt{2 Z_h Z_{H_c}}}\,.
\end{equation}
 
Similar expressions can be obtained for models involving other mirror fermions.  When kinematically accessible, the partial decay width of the Higgs boson into mirror fermions {($\tilde q=\tilde b,\tilde \tau$)} is
\begin{align}\label{eq:decayfermion}
\Gamma (h\to \tilde q \bar{\tilde q})=N_C\frac{(\tilde{\lambda}_{q}^{\rm eff})^2}{8\pi}m_h \left[1-\frac{4m_{\tilde{q}}^2}{m_h^2}\right]^{3/2}\,,
\end{align}
where $N_C$ is the color factor $N_C=3(1)$ for mirror quarks (leptons).  In our numerical analysis,  we will assume for simplicity a ${\mathbb Z}_2$ symmetry between all the Yukawa couplings and their mirror counterpart.  Note, however, that due to the smallness of $m_b$ and $m_\tau$,  the corresponding Yukawa couplings do not have to respect the $\mathbb  Z_2$ symmetry as long as no new naturalness problem is introduced.  The total invisible decay rate into mirror fermions is therefore
\begin{equation}
\Gamma^{\rm tot}_{h\to \tilde f \bar{\tilde f}}=\Gamma (h\to \tilde b \bar{\tilde b})+\Gamma (h\to \tilde \tau\bar{\tilde \tau})\,.
\end{equation}

Concerning the EW gauge sector, after kinematically normalizing the Higgs field, the coupling of the physical Higgs boson to vector bosons is given by
\begin{eqnarray}
\mathcal{L}_{hVV}&=&\frac{h}{\sqrt{Z_h}v}\left[\frac{g^2 v^2}{2}W_{\mu}^+W^{\mu -}+\frac{(g^2+g'^2) v^2}{4}Z_{\mu}Z^{\mu}\right]+ \textrm{mirror terms}\,.
\end{eqnarray}

\subsection{Loop-Induced Decay into Mirror Gluons}
The decay of the Higgs boson into mirror gluons is mediated by loops involving heavy mirror quarks. We estimate such decay process using the effective Yukawa coupling of Eq.~\eqref{eq:YukawaCoupling} for the mirror top quark.    Note that this is  equivalent to defining a dimension-six operator $c_{h\tilde{g}\tilde{g}}|H|^2\tilde{G}^a_{\mu\nu}\tilde{G}^{a,\mu\nu}$ after integrating out the mirror top quark at the scale $m_{\tilde{t}}$ and run it down to the EW scale, We obtain
\begin{align}
\Gamma (h\to \tilde{G}\tilde{G})=\frac{\tilde\lambda_{t}^{\mathrm{eff}\,2} \tilde \alpha_s^2(m_h) m_h^3}{128\pi^3 m_{\tilde{t}}^2}\left|F_{1/2}\left(\frac{4m_{\tilde{t}}^2}{m_h^2}\right)\right|^2{\sqrt{1-\frac{4m_0^2}{m_h^2}}},
\end{align}
where $m_0^2$ is the mass of the mirror glueball  with values in the $12\sim 35$ GeV range~\cite{Curtin:2015fna}.     The loop function of $F_{1/2}$ can be found in Ref.~\cite{Branco:2011iw}.  This channel could also contribute to the Higgs invisible decay if mirror glueball is  stable inside the detector.   For short decay lifetime of the mirror glueball, it could leave a displaced vertex signature which could be observed at future colliders~\cite{Branco:2011iw, Craig:2015pha}.

\section{Global fit and results}\label{sec:globalfit}

The SM Higgs properties have been well measured at the LHC Run-II~\cite{Aad:2019mbh}, and are expected to be constrained with even better precisions at future Higgs factories.
In what follows, we explore the implication of Higgs precision measurements on neutral naturalness models for LHC Run-II~\cite{Aad:2019mbh}, HL-LHC with 3 ab$^{-1}$ integrated luminosity (including  both ATLAS and CMS)~\cite{Cepeda:2019klc}, as well as the CEPC program with 5.6 ab$^{-1}$  integrated luminosity~\cite{Chen:2019pkq,CEPCStudyGroup:2018ghi}.
 
 \begin{figure}
 \centering
\includegraphics[width=0.45\textwidth ]{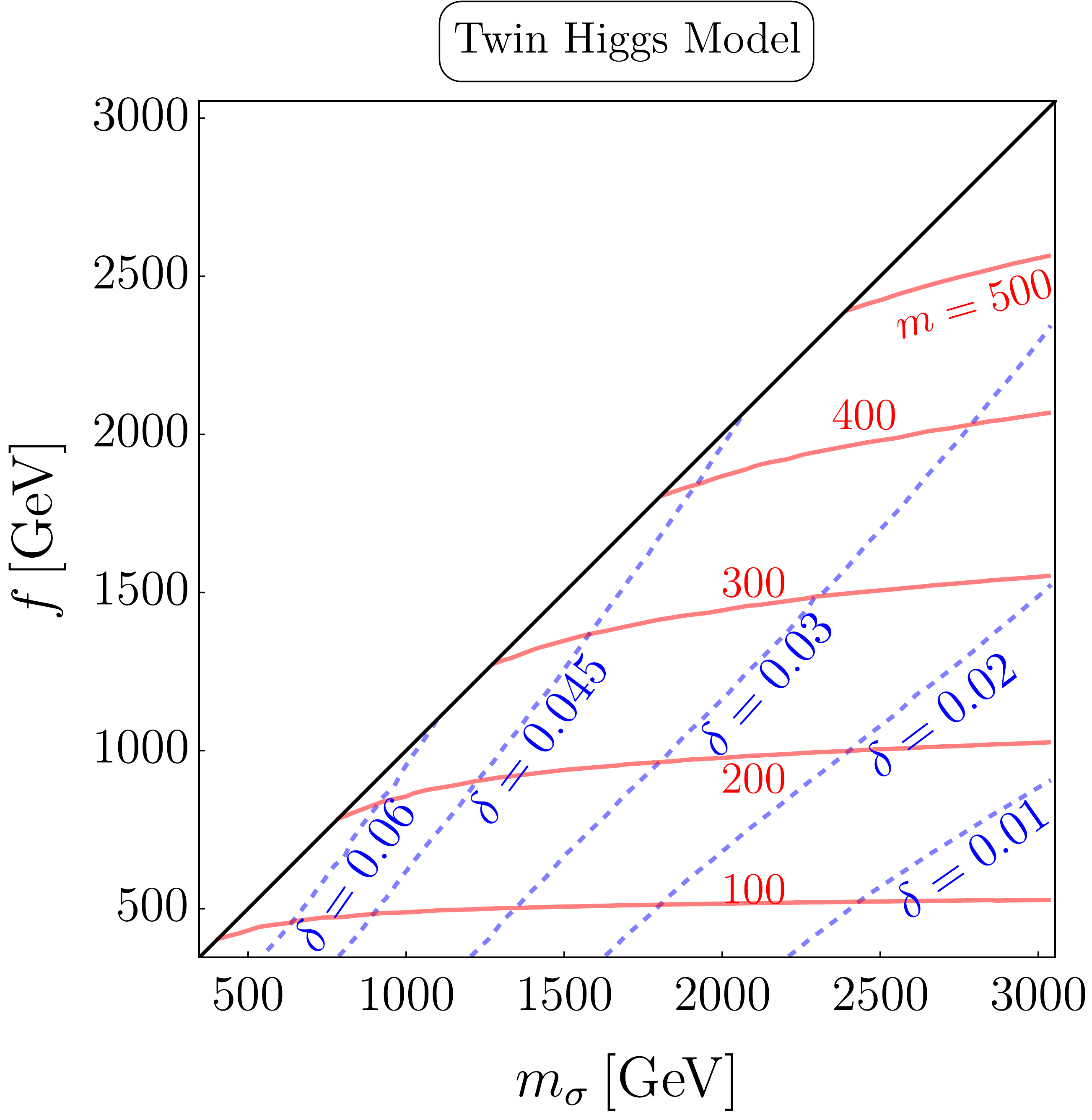}
\includegraphics[width=0.45\textwidth ]{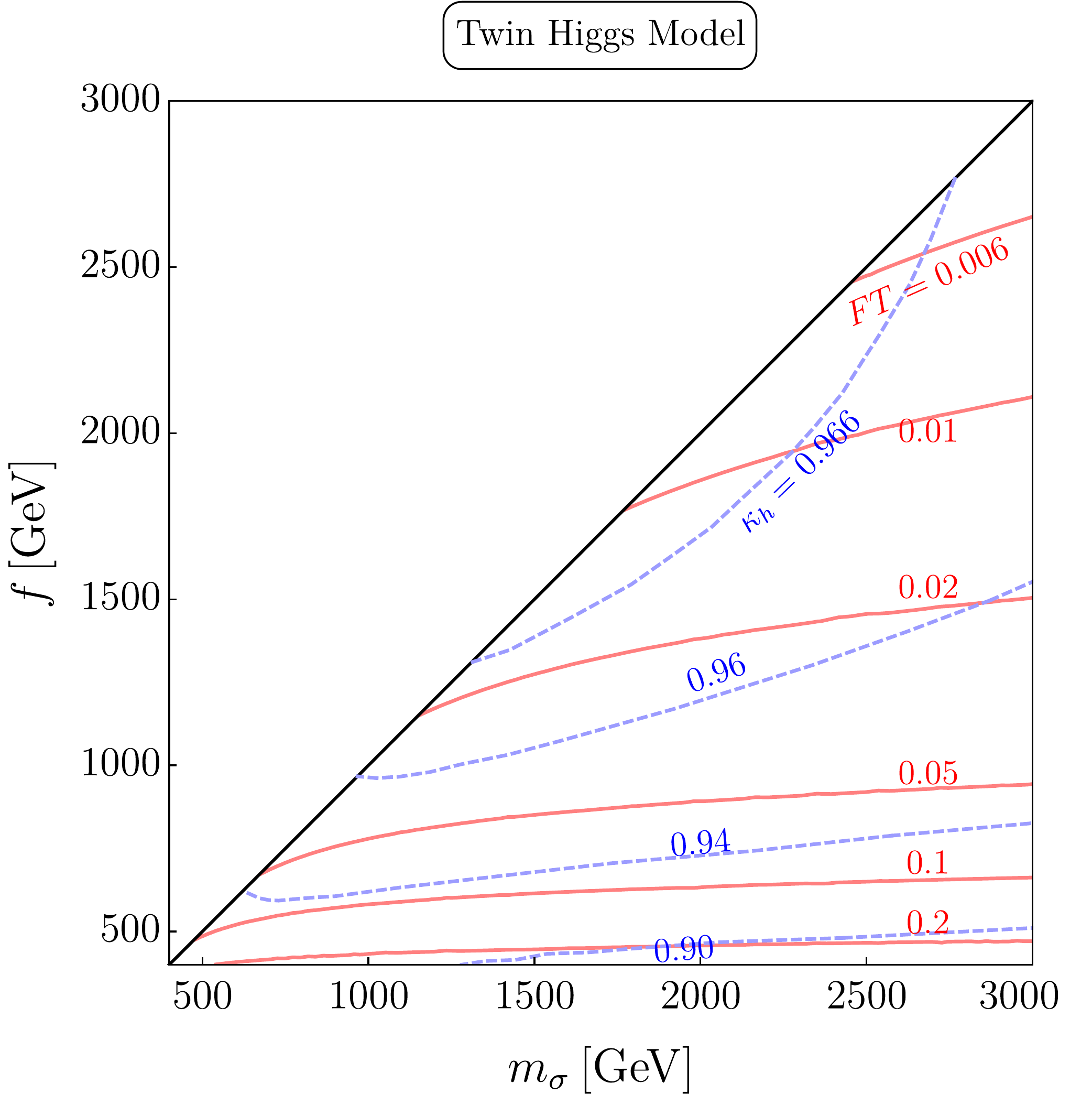}
 \caption{\label{fig:twin_ana} Left panel shows the contours of $m$ (red lines) and $\delta$ (dashed blue lines) in the $m_{\sigma}$ vs. $f$ parameter space of the twin Higgs doublet model. Right panel shows the contours of  fine-tuning level (red lines) and normalized triple Higgs coupling (dashed blue lines) $\kappa_h=\lambda_{hhh}^{BSM}/\lambda_{hhh}^{SM}$.}
\end{figure}
 
 Given the four parameters in the model $(f,\ m_\sigma\ m, \delta)$, two of those ($m$ and $\delta$) can be determined by requiring $m_h=125$ GeV and electroweak vev $v=246$ GeV.   Scanning over the parameter space  shows that solutions can only be obtained for $m_\sigma \geqslant f$. The left panel of Fig.~\ref{fig:twin_ana} shows contours of $m$ (red lines) and $\delta$ (dashed blue lines) in the $m_{\sigma}$ vs. $f$ parameter space, in the case of the twin Higgs model.  The parameter $m$ varies in the range of  $100 - 500$ GeV and $\delta$ varies from  0.01 to 0.06 for $m_\sigma$ and $f$ varying between 400 $-$ 3000 GeV.  
The ratio between $f$ and $m$ is around $4-6$, which does not strongly depend on the mass of the radial mode.  The $\mathbb{Z}_2$ preserving and global symmetry breaking parameter $\delta$ is typically less than $0.1$. 
 
We also estimate the fine-tuning level in the twin Higgs  model.  Following the definition of the fine-tuning in Ref.~\cite{Barbieri:1987fn, Hidalgo:2005js}, we first write $v^2 = v^2(p_1, p_2, \cdots)$ with $p_i$ being the   parameters ($ f, m_\sigma, m, \delta$)  in the model which give rise to the correct Higgs vev, and then define the $\delta v^2 =v^2(p_i+\delta p_i)-v^2(p_i)$ as the change in $v^2$ induced by a change of $\delta p_{i}$ in $p_i$. The fine tuning parameters $\Delta_{p_i}$ associated with $p_i$ and the total fine-tuning ${\textrm{FT}}$ level are defined as
\begin{eqnarray}
	\frac{\delta v^{2}}{v^{2}}=\Delta_{p_{i}} \frac{\delta p_{i}}{p_{i}}, \qquad
	{\textrm{FT}}=\frac{1}{\sqrt{\Sigma_i \Delta_{p_i}^2}},
	\label{eq:ft}
\end{eqnarray}
 with a smaller FT representing a higher fine tuning level. In the right panel of the Fig.~\ref{fig:twin_ana}, we show the fine-tuning level in red contours in the $m_{\sigma}$ vs. $f$ parameter space.  Generally speaking, fine-tuning level is less than 10\% (1\%)   when $f<  600 (2000)$ GeV. Numerically, we also find that   FT is dominated by the parameter $m$ and its contribution $\Delta_{m}$ from~\autoref{eq:ft}. For a fixed value of  $m_\sigma$, a larger $f$ gives rise  a higher fine tuning level, which confirms that $f$ characterizes the misalignment between the electroweak vev $v$ and the spontaneous  global symmetry breaking scale $f$.    Furthermore, we find that the fine tuning level has mild dependence on $m_\sigma$ when $m_\sigma$ is close to $f$ while roughly remain as constant when $m_\sigma \gg f$, due to the decoupling behaviour.

The right panel of Fig.~\ref{fig:twin_ana} also shows the contours of the triple Higgs coupling (normalized to the SM value) $\kappa_{h}=\lambda_{hhh}^{BSM}/\lambda_{hhh}^{SM}$ in dashed blue lines, for the twin Higgs model.   The deviation is typically less than   10\%   over the interested region of parameter space, which agrees with the rough estimation in Ref.~\cite{Agrawal:2019bpm}.  We also evaluated the  amount of fine tuning and the normalized triple Higgs coupling in the cases of ${\rm SO}(5)/{\rm SO}(4)$ and ${\rm SO}(6)/{\rm SO}(5)$. These values are quite similar to the twin Higgs model case.

In order to perform a global fit to the Higgs precision observables, we  construct  the $\chi^2$ distribution using the profile likelihood method,
\begin{eqnarray}
\label{eq:chitot}
\chi^2 &=& \Sigma_i \frac{(\mu_i^{\rm BSM}-\mu_i^{\rm obs})^2}{\sigma_{\mu_i}^2} + 
\frac{(\Sigma_i{\rm Br}^{\rm Inv, BSM}_i-{\rm Br}^{\rm Inv, obs})^2}{(\sigma_{{\rm BR}^{\rm Inv}})^2}\, ,
 \end{eqnarray}
with the sum running over all relevant channels.  
The first sum evaluates the differences between the signal strengths of various Higgs search channels in our model $\mu_i^{\rm{BSM}}=(\sigma\times\textrm{Br}_i)_{\rm{BSM}}/(\sigma\times\textrm{Br}_i)_{\rm{SM}}$ and their observed values  $\mu_i^{\rm obs}$, where $\sigma_{\mu_i}$ is the precision for each process at various colliders.   Usually, the correlations among different $\sigma\times\rm{Br}$ are not provided for  future colliders and are thus assumed to be zero.   
The second term concerns the invisible decay of the Higgs boson into mirror sector, with the prospective limit of  $\rm BR_{\rm Inv} <30\%$ for the current LHC Run-II, ~\cite{Aad:2019mbh}, 4.6\% at  HL-LHC \cite{Cepeda:2019klc}, and 0.3\% at  CEPC \cite{CEPCStudyGroup:2018ghi}.      For future experiments, we assume no deviation from the SM is observed:  $\mu_i^{\rm{obs}}= \mu_i^{\rm{SM}}=1$ and ${\rm Br}^{\rm Inv, obs}=0$.

 \begin{figure}[tbh]
\centering
\includegraphics[width=0.48\textwidth,clip=true, trim = 47mm 150mm 55mm 20mm ]{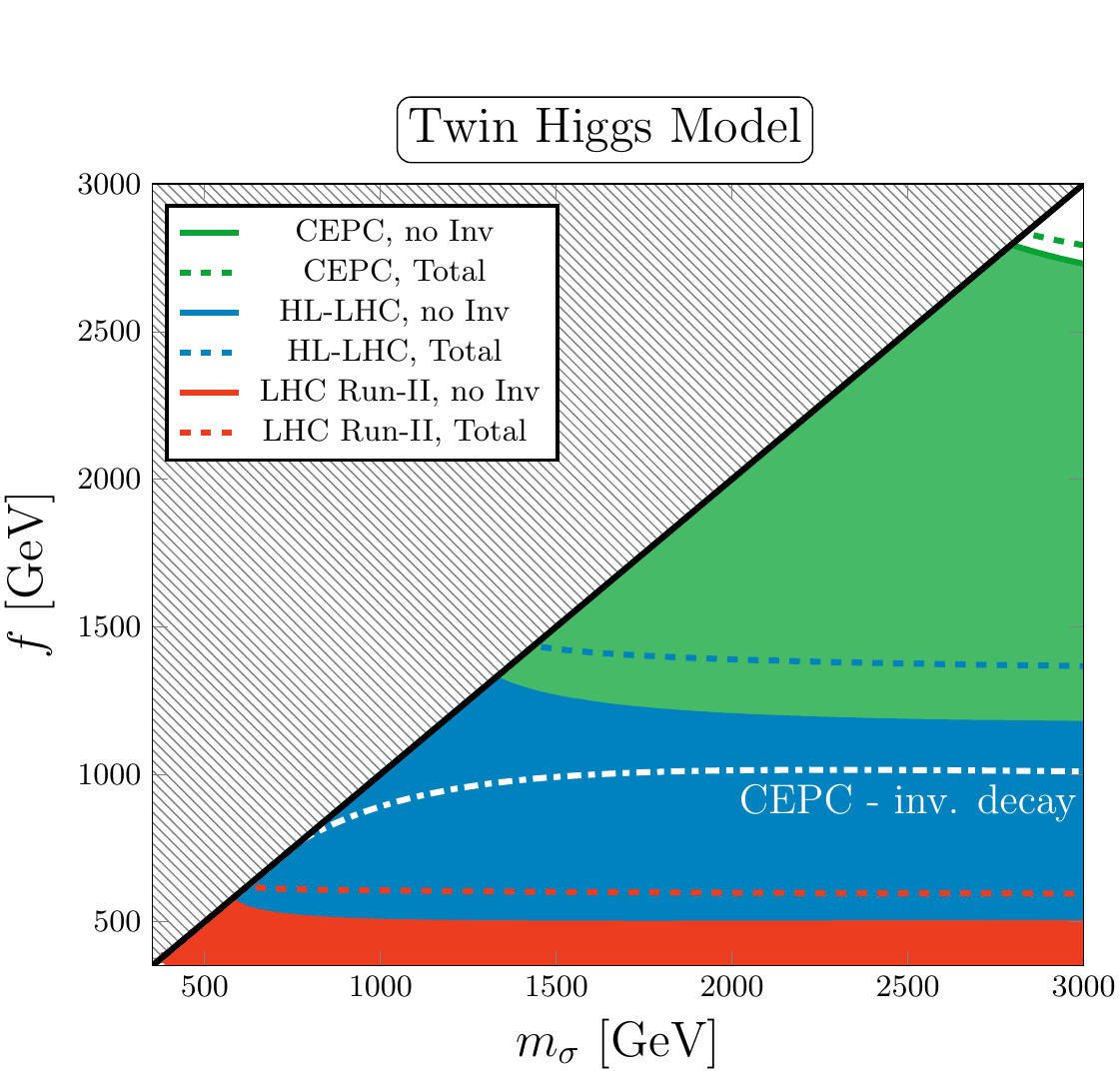}
\includegraphics[width=0.48\textwidth,clip=true, trim = 47mm 150mm 55mm 20mm ]{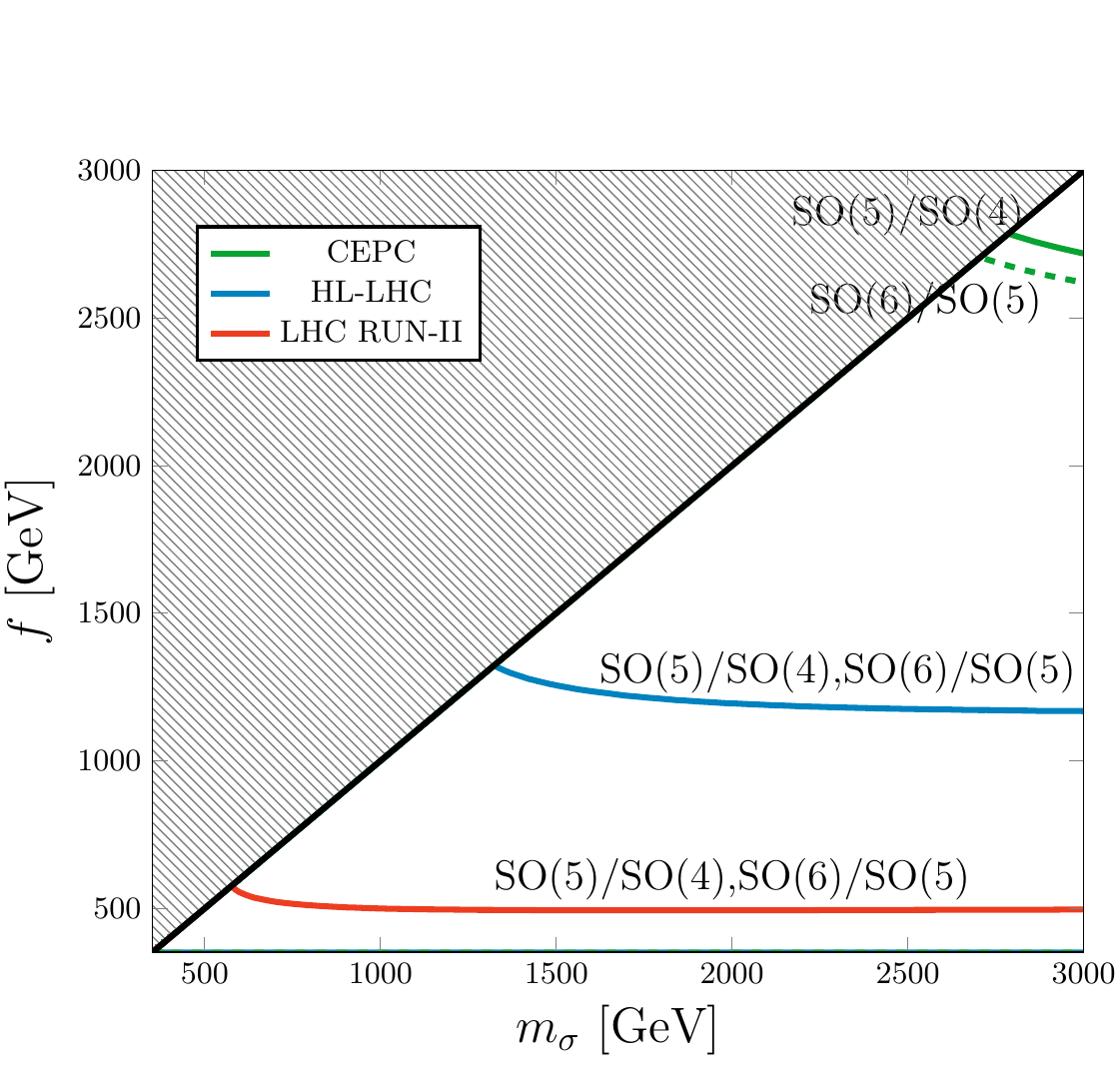}
 \caption{\label{fig:fit_twin} 95\% C.L. excluded region  in the $m_\sigma - f$ plane for the twin Higgs double model (left panel) and for the other two cosets ${\rm SO}(5)/{\rm SO}(4)$ and ${\rm SO}(6)/{\rm SO}(5)$ (right panel),  for the Higgs precision measurements at the CEPC (green), HL-LHC (blue) and the current LHC Run-II (red). In the left panel, shaded regions stand for constraints including only Higgs visible decay channels whereas colored lines  include Higgs invisible decay in addition.  
The white dash-dotted line represents the constraints on the Higgs invisible decay alone from the CEPC.   
In the right panel, solid and dashed lines indicate the region for ${\rm SO}(5)/{\rm SO}(4)$ and ${\rm SO}(6)/{\rm SO}(5)$, respectively. 
}

\end{figure}

The left panel of Fig.~\ref{fig:fit_twin} shows the 95\% C.L. region in the $m_\sigma - f$ plane for the twin Higgs  model, which corresponds to  $\Delta\chi^2=\chi^2-\chi_{\rm{min}}^2 \leqslant 5.99$. The green region, covering most of the parameter space up to $f \sim 2700$ GeV, is expected to be excluded by Higgs precision measurements at the CEPC if no deviation from the SM is observed. The dashed green line includes the Higgs invisible decay in addition, which extends the reach for about 100 GeV.  The reach of the Higgs invisible decay alone via the CEPC measurements is given by the white dashed line. The constraints on $f$ from invisible decay get stronger for larger  $m_\sigma$, which reaches $\sim 1$ TeV for $m_{\sigma}\gtrsim 1.8$ TeV.  The blue and red regions represent similar limits with  the HL-LHC and the current LHC Run-II precision measurements, up to $f= 1300, 500$ GeV respectively. The constraint gets weaker with larger $m_\sigma$ and stays flat, due to the decoupling of the radial mode.   Although the precision of   Higgs invisible decay at the LHC Run-II and HL-LHC is relatively worse comparing to Higgs factories,   the reach  including the Higgs invisible decay channel, represented by the dashed lines of the corresponding color, are about $\mathcal O(100)$ and $\mathcal O(200)$ GeV larger than the ones involving visible decays only. 

We notice that the exclusion limits are not sensitive to the mass of the radial mode, except when its value is close to the spontaneous global symmetry breaking vev, $f$. This can be understood from the expression of the Higgs couplings to the SM particles normalized to the SM values,  which have the form of $\kappa\simeq 1-c_H\frac{v^2}{2f^2}=1-\frac{v^2}{4f^2}-(1-2\delta)\frac{2v^2 m^2}{f^2 m_\sigma^2}$.   The last term decouples for large $m_\sigma$, while contributing negatively to $\kappa$ for $m_\sigma \sim f$.

In Ref.~\cite{Chacko:2017xpd}, the authors also consider a mirror twin Higgs model with radial mode, where they find the   Higgs coupling measurements can exclude a twin top mass up to 670~GeV at the HL-LHC. This can be translate into $f$ up to 950~GeV in our case.  Similar conclusion is also found in Ref.~\cite{Burdman:2014zta} where no radial mode is included.  However, in both studies,   the RG running contributions are ignored.  We also perform a global fit to all the Higgs decay channels instead of just SM visible channels.

The right panel of Fig.~\ref{fig:fit_twin} shows the global fit results for the other two cosets ${\rm SO}(5)/{\rm SO}(4)$ (solid lines) and ${\rm SO}(6)/{\rm SO}(5)$ (dashed lines).  The region beneath the lines are excluded.  For these two cosets, since there is no Higgs invisible decay into heavy mirror fermions, the effect of including the Higgs invisible decay to the mirror glueball  are too small to be significant.  For the  HL-LHC and LHC Run-II results, the limits for the two cosets are degenerate.

\section{Conclusion}
\label{sec:conclusion}

 A class of neutral naturalness models presents an universal structure that the SM Higgs is the pseudo-Goldstone boson associated with the spontaneous breaking of some global symmetry.  Although such symmetry is explicitly broken by its Yukawa coupling with fermions, the Higgs mass is protected from large quadratic radiative contributions via a discrete mirror symmetry.  The mirror partner of the SM top quark is charged under a mirror QCD while being neutral under the SM strong interactions.    In this paper, we investigated the Higgs sector in several neutral naturalness models using an effective field theory description at different scales.  At the UV scale, the Lagrangian of the Higgs sector is parameterized using the pseudo-Goldstone Higgs chiral Lagrangian with ${\rm SO}(N)/{\rm SO}(N-1)$ coset including the radial mode.  The ${\rm SO}(N)/{\rm SO}(N-1)$ coset Lagrangian uniformly describes several neutral naturalness models, including the twin Higgs   model, the minimal neutral naturalness and the trigeometric Higgs models. Below the scale of spontaneous global symmetry breaking with  the radial mode integrated out,  the chiral Lagrangian is matched to the dimension-six operators in the Standard Model effective field theory.  
Below the electroweak symmetry breaking scale, the Lagrangian is matched to the Higgs couplings with generic Lorentz structures, from which the Higgs couplings in various models could be extracted and explored.  

In order to obtain the realistic Higgs mass, the above tree-level matching procedure is not enough.  Indeed, for a given effective field theory valid at some UV physics scale $m_\sigma$, the running of the different parameters from the scale $m_\sigma$ to the electroweak scale in the presence of dimension-six operators can lead to large variations of the electroweak vev and of the Higgs boson mass as compared to the values one would obtain by simply minimizing the UV effective potential. 
In order to take those effects into accounts, we proceeded in the following way: $(i)$ at the scale $m_\sigma$, we integrated out the radial mode at tree level and defined our UV effective field theory. $(ii)$ We computed the running of the effective Higgs potential from the scale $m_\sigma$ to the electroweak scale arising from loops of scalars, fermions and gauge bosons within this EFT.
$(iii)$ After electroweak symmetry breaking and proper kinetic normalization of the Higgs field, we obtained the values of the Higgs vev and physical Higgs mass at low energy as a function of the UV parameters of our model.

Given the effective Higgs couplings at the electroweak scale, we calculated the signal strength for various Higgs decay channels, and performed a global fit to the Higgs precision measurements at current and future colliders.   We found that   only region of the parameter space with  $m_\sigma \geqslant f$ can accommodate the correct Higgs mass $m_h$ and vacuum expectation value $v$ for all three  scenarios.  The 95\% C.L. reach on  parameter $f$ are shown in Table~\ref{tab:exclu_f} for various current and future colliders, if no deviation of the SM Higgs couplings is observed.  Typically, invisible decays could improve the $f$ constraints  by about 100 GeV.   The triple Higgs couplings can be modified at 5$-$10\% level, which could hopefully be probed at future high-energy colliders~\cite{Benedikt:2018csr}.  The level of fine-tuning is dominated by the parameter $m$, which is less than 10\% (1\%)  when $f< 600~  (2000)$ GeV.

\begin{table}
\center
\begin{tabular}{c|c|c|c}
\hline 
\hline
$f$ (GeV) & LHC Run-II & HL-LHC & CEPC  \\
 \hline 
twin Higgs & 500 & 1300 &  2700      \\
 ${\rm SO}(5)/{\rm SO}(4)$ & 500 & 1100 & 2600     \\
${\rm SO}(6)/{\rm SO}(5)$ & 500 & 1100 & 2700  \\
\hline 
\hline
\end{tabular}
\caption{95\% C.L. constraints on the spontaneous global symmetry breaking scale $f$ at current and future colliders, assuming no deviation from the SM predictions of the Higgs couplings is observed.}
\label{tab:exclu_f}
\end{table}

While the precision Higgs measurements are mostly sensitive to the the scale of the spontaneous global symmetry breaking $f$, it is complementary to other experimental probes of the neutral naturalness models. In the future,  experiments detecting long lived particle~\cite{Lubatti:2019vkf,Alidra:2020thg}  could probe the Higgs exotic decay
channels through the displaced vertices signatures of the mirror glueball.    
Furthermore,  heavy radial mode can be directly produced at a future 100 TeV $pp$ collider.  The non-resonant and resonant  di-Higgs searches at the HL-LHC and future 100 TeV colliders could help us to explore the scalar potential of the Higgs sector.  Combining all the direct and indirect searches allows us to gain insight into the dark sector of neutral naturalness models.

\appendix
\section{Parameterization of ${\mathcal H}$ for each model}
\label{appendixA}
Different coset structures have the multiplet $\cal H$ parameterized in the following forms under unitary gauge:  
\begin{eqnarray}
&&{\rm twin\ Higgs\ ({\rm SU}(4)/{\rm SU}(3)):}{\cal H}=(f+\sigma/\sqrt{2})(\frac{H^T}{|H|}\sin{\frac{|H|}{f}},0,\cos(\frac{|H|}{f}) )^T,\\
&&{\rm twin\ Higgs\ }({\rm SO}(8)/{\rm SO}(7)): {\cal H}=(f+\sigma/\sqrt{2})(\frac{\Pi_{i=1-4}}{|\Pi|}\sin\frac{|\Pi|}{f},0,0,0,\cos(\frac{|\Pi|}{f}) )^T,\\
&&{\rm SO}(5)/{\rm SO}(4): {\cal H}=(f+\sigma/\sqrt{2})(\frac{\Pi_{i=1-4}}{|\Pi|}\sin\frac{|\Pi|}{f},\cos(\frac{|\Pi|}{f}) )^T,\\
&&{\rm SO}(6)/{\rm SO}(5): {\cal H}=(f+\sigma/\sqrt{2})(\frac{\Pi_{i=1-4}}{|\Pi|}\sin\frac{|\Pi|}{f},0,\cos(\frac{|\Pi|}{f}) )^T,
\end{eqnarray}
where in the twin Higgs ${\rm SU}(4)/{\rm SU}(3)$, the $H$ represents the ordinary SM Higgs doublet, while in the twin Higgs ${\rm SO}(8)/{\rm SO}(7)$, neutral naturalness models ${\rm SO}(5)/{\rm SO}(4)$ and ${\rm SO}(6)/{\rm SO}(5)$, $\Pi_i$ and Higgs doublet have the following relation:
\begin{eqnarray}
H=\begin{pmatrix}
\Pi_2+i\Pi_1\\
\Pi_4-i\Pi_3
\end{pmatrix}.
\end{eqnarray} 
In addition,   we summarize the Goldstone matrix ${\cal U}$ in   Table~\ref{tab:gs_models}.
\begin{table}[ht]\center
\begin{tabular}{c|c}
\hline 
\hline
 coset & ${\cal U}$ matrix    \\
 \hline 
twin Higgs ${\rm SU}(4)/{\rm SU}(3)$&   $\left(
\begin{array}{cc|c|c}
    \multicolumn{2}{c|}{\multirow{2}{*}{${1}-\left(1-\cos{\frac{|H|}{f}}\right)\frac{HH^\dagger}{|H^2|}$}} & 0 &  \multirow{2}{*}{$\frac{H}{|H|}\sin{\frac{|H|}{f}}$} \\
    & & 0 &\\
    \hline
    0 & 0 & 0 & 0\\
    \hline
    \multicolumn{2}{c|}{  
    \begin{matrix}
  		-\frac{H^\dagger}{|H|}\sin{\frac{|H|}{f}}
  \end{matrix}}   & 0& \cos{\frac{|H|}{f}} 
\end{array}
\right) $   \\
\hline
twin Higgs ${\rm SO}(8)/{\rm SO}(7)$&  $\left(
\begin{array}{cc|c|c}
    \multicolumn{2}{c|}{\multirow{2}{*}{${1}-\left(1-\cos{\frac{|\Pi|}{f}}\right)\frac{\Pi_i\Pi^\dagger_j}{|\Pi^2|}$}} & \multirow{2}{*}{$0_{4\times 3}$} &  \multirow{2}{*}{$\frac{\Pi_i}{|\Pi|}\sin{\frac{|\Pi|}{f}}$} \\ 
    & & &\\
    \hline
    \multicolumn{2}{c|}{0_{3\times 4}} & 1_{3\times 3} & 0_{3\times 1}\\
    \hline
    \multicolumn{2}{c|}{  
    \begin{matrix}
  		-\frac{\Pi^\dagger_j}{|\Pi|}\sin{\frac{|\Pi|}{f}}
  \end{matrix}}   & 0& \cos{\frac{|\Pi|}{f}} 
\end{array}
\right) $\\
\hline
 ${\rm SO}(5)/{\rm SO}(4)$&   $\left(
\begin{array}{cc|c}
    \multicolumn{2}{c|}{\multirow{2}{*}{${1}-\left(1-\cos{\frac{|\Pi|}{f}}\right)\frac{\Pi_i\Pi^\dagger_j}{|\Pi|^2}$}} &   \multirow{2}{*}{$\frac{\Pi_i}{|\Pi|}\sin{\frac{|\Pi|}{f}}$} \\
    &  & \\
    \hline
    \multicolumn{2}{c|}{  
    \begin{matrix}
  		-\frac{\Pi^\dagger_j}{|\Pi|}\sin{\frac{|\Pi|}{f}}
  \end{matrix}}   &  \cos{\frac{|\Pi|}{f}} 
\end{array}
\right) $   \\
\hline
${\rm SO}(6)/{\rm SO}(5)$&   $\left(
\begin{array}{cc|c|c}
    \multicolumn{2}{c|}{\multirow{2}{*}{${1}-\left(1-\cos{\frac{|\Pi|}{f}}\right)\frac{\Pi_i\Pi^\dagger_j}{|\Pi^2|}$}} & \multirow{2}{*}{$0_{4\times 1}$} &  \multirow{2}{*}{$\frac{\Pi_i}{|\Pi|}\sin{\frac{|\Pi|}{f}}$} \\ 
    & & &\\
    \hline
    \multicolumn{2}{c|}{0_{1\times 4}} & 1 & 0\\
    \hline
    \multicolumn{2}{c|}{  
    \begin{matrix}
  		-\frac{\Pi^\dagger_j}{|\Pi|}\sin{\frac{|\Pi|}{f}}
  \end{matrix}}   & 0& \cos{\frac{|\Pi|}{f}} 
\end{array}
\right) $\\
\hline 
\hline
\end{tabular}
\caption{Goldstone Matrices for different cosets, where we have already set the Goldstones not in the SM Higgs $H$ to zero under unitary gauge.  }
\label{tab:gs_models}
\end{table}

Two matrices $\bold m^2$ and ${\boldsymbol \delta}$ are defined as:
\begin{eqnarray}
\bold m^2=
\begin{pmatrix}
m^2{1}_{n\times n}& 0\\
0 & -m^2{1}_{p\times p}
\end{pmatrix},
\quad
\boldsymbol \delta=
\begin{pmatrix}
\sqrt{\delta} {1}_{n\times n}& 0\\
0 & -i\sqrt{\delta}{1}_{p\times p}
\end{pmatrix},
\end{eqnarray}
where the $n$ and $p$ depend on the different coset structures. Based on representations we specify above, The corresponding    $(n,p)$ of the representations   are summarized in the Table~\ref{tab:np_models_2}.

\begin{table}\center
\begin{tabular}{c|cc}
\hline 
\hline
 coset & n & p    \\
 \hline 
twin Higgs ${\rm SU}(4)/{\rm SU}(3)$ &  2 & 2     \\
twin Higgs ${\rm SO}(8)/{\rm SO}(7)$&  4 & 4     \\
 ${\rm SO}(5)/{\rm SO}(4)$& 4 & 1    \\
${\rm SO}(6)/{\rm SO}(5)$ & 4 & 2  \\
\hline 
\hline
\end{tabular}
\caption{$n$ and $p$ for different coset structure.  }
\label{tab:np_models_2}
\end{table}

The covariant derivative $D_\mu$ is defined as:
\begin{eqnarray}
D_{\mu}&=&\begin{pmatrix}
D_{\mu}^A & 0_{n\times p}\\
0_{p\times n} & D_{\mu}^B
\end{pmatrix},
\end{eqnarray}
where the concrete forms of $D^A_{\mu}$, and $D^B_{\mu}$ for different coset structure are summarized in the Table~\ref{tab:der_models}. 

\begin{table}[ht]\center
\begin{tabular}{c|cc}
\hline 
\hline
 coset & $D_{\mu}^A$ & $D_{\mu}^B$    \\
 \hline 
twin Higgs ${\rm SU}(4)/{\rm SU}(3)$ &  $\partial_{\mu}{1}_{2\times 2}-ig\frac{\sigma^\alpha}{2}W_\mu^\alpha- i\frac{g'}{2}B_\mu$ & $\partial_{\mu}{1}_{2\times 2}-i\tilde{g}\frac{\sigma^\alpha}{2}\tilde{W}_\mu^\alpha- i\frac{\tilde{g}'}{2}\tilde{B}_\mu$     \\
twin Higgs ${\rm SO}(8)/{\rm SO}(7)$ &  $\partial_{\mu}{1}_{4\times 4}-igt_L^\alpha W_\mu^\alpha- ig' t_R^3B_\mu$ & $\partial_{\mu}{1}_{4\times 4}-i\tilde{g}t_L^\alpha \tilde{W}_\mu^\alpha- i\tilde{g}' t_R^3\tilde{B}_\mu$     \\
 ${\rm SO}(5)/{\rm SO}(4)$& $\partial_{\mu}{1}_{4\times 4}-igt_L^\alpha W_\mu^\alpha- ig't_R^3 B_\mu$ & $\partial_{\mu}{1}_{1\times 1}$     \\
${\rm SO}(6)/{\rm SO}(5)$ & $\partial_{\mu}{1}_{4\times 4}-igt_L^\alpha W_\mu^\alpha- ig't_R^3 B_\mu$ & $\partial_{\mu}{1}_{2\times 2}-i\tilde{g}_1\frac{\sigma^{2}}{\sqrt{2}}\tilde{B}'_\mu$   \\
\hline 
\hline
\end{tabular}
\caption{Covariant derivative in different models.    }
\label{tab:der_models}
\end{table}
The $t_L$ and $t_R$ in the table are the $4\times4$ matrix defined by:
\begin{eqnarray}
(t^\alpha_L)_{ij}=\frac{1}{2\sqrt{2}}{\rm Tr}[\bar{\sigma}^\dagger_i\sigma^\alpha \bar{\sigma}_j],\\
(t^\alpha_R)_{ij}=\frac{1}{2\sqrt{2}}{\rm Tr}[\bar{\sigma}_i\sigma^\alpha \bar{\sigma}^\dagger_j],
\end{eqnarray}
with $i,j$ ranging from 1 to 4.    $\sigma^{\alpha =1,2,3}$ is the ordinary Pauli matrices and   $\bar{\sigma}$ is defined by:
\begin{eqnarray}
\bar{\sigma}=(i\sigma^\alpha,{1}_{2\times 2}).
\end{eqnarray}

\begin{acknowledgments}

We thank J\'er\'emie Qu\'evillon and Justin Lieffers for useful discussions.
HLL also thanks Ming-Lei Xiao for helpful discussion.
LH, HS and SS are  supported by the Department of Energy under Grant DE-SC0009913.
HLL and JHY are supported by the National Science Foundation of China (NSFC) under Grants No. 11875003.
JHY is also supported by the National Science Foundation of China (NSFC) under Grants No. 11947302. 
WS is supported by the Australian Research Council Discovery Project DP180102209.
The work of LH has been partially performed during the workshop ``Dark Matter as a Portal to New Physics" supported by Asia Pacific Center for Theoretical Physics and by the Institut Pascal at Universit\'e Paris-Saclay with the support of the P2I and SPU research departments and
the P2IO Laboratory of Excellence (program ``Investissements d'avenir"
ANR-11-IDEX-0003-01 Paris-Saclay and ANR-10-LABX-0038), as well as the
IPhT.

 \end{acknowledgments}

\bibliographystyle{JHEP}

\bibliography{twineft}
\end{document}